\documentclass[reprint,superscriptaddress,amsmath,amssymb,pra]{revtex4-2}
\usepackage{bm}
\usepackage{graphicx}
\usepackage{color}
\usepackage{txfonts}

\newcommand{\p}{\mathrm{p}}
\renewcommand{\i}{\mathrm{i}}
\newcommand{\s}{\mathrm{s}}
\newcommand{\sinc}{\mathrm{sinc}}
\newcommand{\e}{\mathrm{e}}
\newcommand{\vac}{\text{vac}}
\newcommand{\twin}{\text{twin}}

\newcommand{\SE}{\mathrm{SE}}

\newcommand{\rephasing}{\mathrm{(r)}}
\newcommand{\nonrephasing}{\mathrm{(nr)}}

\newcommand{\adagger}{\hat{a}^\dagger}

\newcommand{\tr}{\mathrm{tr}}

\renewcommand{\Re}{\mathrm{Re}}

\newcommand{\E}{\hat{E}}

\begin{document}

\title{Two-dimensional fluorescence spectroscopy with quantum entangled photons and time- and frequency-resolved two-photon coincidence detection}

\author{Yuta Fujihashi}
\affiliation{Department of Engineering Science, The University of Electro-Communications, Chofu 182-8585, Japan}

\author{Ozora Iso}
\affiliation{Department of Engineering Science, The University of Electro-Communications, Chofu 182-8585, Japan}

\author{Ryosuke Shimizu}
\affiliation{Department of Engineering Science, The University of Electro-Communications, Chofu 182-8585, Japan}
\affiliation{Institute for Advanced Science, The University of Electro-Communications, Chofu 182-8585, Japan}

\author{Akihito Ishizaki}
\affiliation{Department of Chemistry, The University of Tokyo, Tokyo 113-0033, Japan}
\affiliation{Institute for Molecular Science, National Institutes of Natural Sciences, Okazaki 444-8585, Japan}

\begin{abstract}
Recent theoretical studies in quantum spectroscopy have emphasized the potential of non-classical correlations in entangled photon pairs for selectively targeting specific nonlinear optical processes in nonlinear optical responses. However, because of the extremely low intensity of the nonlinear optical signal generated by irradiating molecules with entangled photon pairs, time-resolved spectroscopic measurements using entangled photons have yet to be experimentally implemented. In this paper, we theoretically propose a quantum spectroscopy measurement employing a time-resolved fluorescence approach that aligns with the capabilities of current photon detection technologies. The proposed quantum spectroscopy affords two remarkable advantages over conventional two-dimensional electronic spectroscopy. First, it enables the acquisition of two-dimensional spectra without requiring control over multiple pulsed lasers. Second, it reduces the complexity of the spectra because the spectroscopic signal is contingent upon the nonlinear optical process of spontaneous emission. These advantages are similar to those achieved in a previous study [Fujihashi \textit{et al.}, J. Chem. Phys. \textbf{160}, 104201 (2024)]. However, our approach achieves sufficient signal intensities that can be readily detected using existing photon detection technologies, thereby rendering it a practicable. Our findings will potentially facilitate the first experimental real-time observation of dynamic processes in molecular systems using quantum entangled photon pairs.
\end{abstract}

\maketitle

\section{Introduction}
The potential of quantum light to serve as a valuable resource for advancing innovative measurement techniques in spectroscopy has garnered increasing attention \cite{Mukamel:2020ej,Georgiades:1995dd,Georgiades:1997at,Fei:1997es,Lee:2006id,Schlawin:2024tw,Pandya:2024to,Oka:2018two,Munoz:2021qu,Matsuzaki:2022su,Matsuzaki:2024su,Kalashnikov:2016cl,Mukai:2022qu,Tashima:2024ul,Albarelli:2023fu,Khan:2024do,Darsheshdar:2024ro,Rodriguez:2024pe}. For instance, entangled photons have the potential to enable significant advances in the scaling of two-photon absorption \cite{Georgiades:1995dd,Georgiades:1997at,Fei:1997es,Lee:2006id,Schlawin:2024tw,Pandya:2024to}, subshot noise absorption spectroscopy \cite{Matsuzaki:2022su,Matsuzaki:2024su}, and infrared spectroscopy with visible light \cite{Kalashnikov:2016cl,Mukai:2022qu,Tashima:2024ul}. Given these promising capabilities, researchers have begun exploring the feasibility of employing entangled photons in time-resolved spectroscopy, including coherent multidimensional optical spectroscopy \cite{Roslyak:2009mu,Raymer:2013kj,Schlawin:2016er,Zhang:2022en,Ko:2023em,Ishizaki:2020jl,Chen:2022mo,Gu:2023ph,Fujihashi:2023pr,Dorfman:2014bn,Asban:2021di,Kizmann:2023qu,Yadalam:2023qu,Fujihashi:2024pa,Kim2024:pr,Yadalam:2024ul,Fan:2024en,Jadoun:2024ra}. For example, coincidence detection of entangled photon pairs has been shown to enhance the signal-to-noise ratio of pump-probe spectroscopy \cite{Schlawin:2016er}. The non-classical photon correlation between entangled photons has the potential to enable time-resolved spectroscopy with monochromatic pumping \cite{Ishizaki:2020jl,Chen:2022mo,Gu:2023ph,Fujihashi:2023pr} and to selectively target specific nonlinear optical processes in a nonlinear optical response \cite{Dorfman:2014bn,Asban:2021di,Kizmann:2023qu,Yadalam:2023qu,Fujihashi:2024pa,Kim2024:pr,Yadalam:2024ul}.

The nonlinear optical susceptibility of molecules is typically observed to exhibit relatively small magnitudes. Even with the use of a quasi-phase-matched crystal, the conversion efficiency of spontaneous parametric down-conversion (SPDC) remains within the range of $10^{-6}$ to $10^{-10}$. Consequently, the intensity of the nonlinear optical signal generated by irradiating molecules with entangled photon pairs is extremely low \cite{Landes:2021ex,Parzuchowski:2021se,Raymer:2021la}. As the previously proposed time-resolved quantum spectroscopy measurement \cite{Ishizaki:2020jl,Fujihashi:2024pa} relies on this two-photon irradiation, the issue of low signal intensity remains unresolved. Consequently, experimental implementations of time-resolved spectroscopic measurements with entangled photons have yet to be realized.

An alternative approach to overcoming the constraints associated with nonlinear optical signals is the use of entangled photons in time-resolved fluorescence measurements \cite{Scarcelli:2008en,Fujihashi:2020ep,Harper:2023en,Eshun:2023fl,Gabler:2025be,Li:2023si}. Time-resolved fluorescence spectroscopy facilitates the acquisition of data pertaining to third-order nonlinear optical responses by detecting fluorescence signals from the excited states of molecules induced by light irradiation. This method requires irradiating only one of the entangled photon pairs onto the molecules \cite{Fujihashi:2020ep}; this ensures that the signal intensities are sufficiently strong to be detected using current photon detection technology. Researchers have successfully conducted fluorescence lifetime measurements with monochromatic pumping \cite{Harper:2023en,Eshun:2023fl,Gabler:2025be} by leveraging the non-classical correlations of entangled photon pairs. Additionally, single-photon absorption events in photosynthetic complexes have been observed \cite{Li:2023si}. 

However, performing time-resolved and frequency-resolved measurements based on fluorescence measurements with entangled photons presents the challenge of prolonged measurement times. While charge-coupled devices (CCDs) have been employed for absorption measurements at the single-photon level \cite{Kalashnikov:2016cl,Matsuzaki:2022su,Matsuzaki:2024su}, their frame rates impose significant limitations on the permissible number of events. Consequently, CCDs cannot perform time-stamping measurements, leading to extended measurement times when acquiring time-resolved spectra. Recently, a photon detection technique using a delay-line anode single-photon detector (DLD) has been developed \cite{Iso:2024ca}. 
The DLD incorporates a photocathode, microchannel plate, and delay-line anode. Photoelectron signals guided by the microchannel plate provide information on the photon arrival time at the DLD with a time resolution of several hundred picoseconds. Furthermore, the DLD can reconstruct photon arrival positions by analyzing the time difference between signals generated at both ends of the meander-shaped delay-line anode. Thus, combining the DLD with a spectrometer allows single-photon-level time-resolved spectral measurements without the need for frequency scanning. Moreover, the absence of frame rate limitations significantly reduces measurement times. For instance, the two-dimensional frequency distribution of entangled photon pairs generated from a CuCl semiconductor single crystal can be measured within a few minutes---a process that is thousands of times faster than conventional methods. As a result, this technology enables the development of a time-resolved spectroscopy method with entangled photons, capable of simultaneously measuring frequency and time information within a practical measurement duration.

\begin{figure}
    \centering
    \includegraphics{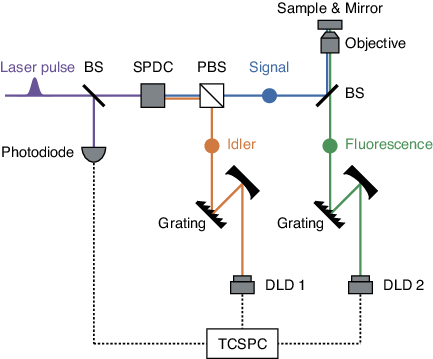}
    \caption{Schematic of the proposed quantum spectroscopy. A pulsed laser pumps a type-II PDC crystal, generating degenerate entangled twin photons. A beam sampler (BS) partially reflects the pump source to a photodiode as a reference signal, which initiates the timer of a time-correlated single-photon counting (TCSPC) device. After passing through the crystal, the photons are sorted by a polarizing beam splitter (PBS): The idler photon is dispersed by a spectrometer and detected by a DLD (DLD~1), while the signal photon passes through the PBS and excites a molecule in a microscope setup. Similarly, the fluorescence photon from the molecule is dispersed and detected by another DLD (DLD~2). The TCSPC device measures time correlation between idler photons and fluorescence photons, both of which pass through a spectrometer and then a DLD before being output. This measurement process is repeated, and the coincidence signal is integrated to obtain the time-resolved fluorescence spectrum as a function of $\bar{\omega}_\mathrm{i}$, $\bar{\omega}_\mathrm{F}$, $\bar{t}_\mathrm{i}$, and $\bar{t}_\mathrm{F}$. Here, $\bar\omega_a$ and $\bar{t}_a$ are the frequency and arrival time of photon~$a$ recorded at the DLD, respectively.
    }
    \label{fig1}
\end{figure}

In this paper, we propose a quantum spectroscopy measurement utilizing a time-resolved fluorescence approach that is compatible with current photon detection technologies. As shown in Fig.~\ref{fig1}, the proposed quantum spectroscopy measurement can obtain time-resolved spectra, including two-dimensional electronic spectra (2DES) \cite{Brixner:2005wu,Tekavec:2007ij,Lott:2011kd,Agathangelou:2021ph}, without requiring control over multiple pulsed lasers. Furthermore, we present a theoretical framework demonstrating that the complexity of the spectra can be reduced because the spectroscopic signal is contingent upon the nonlinear optical process of spontaneous emission. These advantages align with those of previously proposed spectroscopic methods \cite{Ishizaki:2020jl,Fujihashi:2024pa}. However, this approach achieves sufficient signal intensities that can be readily detected with existing photon detection technologies, rendering it a practical method.

\section{Quantum states of entangled twin}
For simplicity, we assumed the following: (1) the weak down-conversion regime, (2) the degenerate type-II SPDC, (3) the SPDC process is perfectly phase-matched at the central frequency $\omega_\mathrm{p}$, and the group velocity dispersion through the nonlinear medium is negligible \cite{Keller:1997hj}, (4) the symmetric group-velocity matching, where the sum of the inverse group velocities of the signal and idler photons is equal to twice the pump inverse group velocity \cite{Ansari:2018ta,Gatti:2018he}, and (5) the impulsive pump limit. On the basis of these assumptions, the quantum state of the generated twin photons is expressed as 
\begin{align}
	\lvert \psi_\twin \rangle
	=
    \zeta
	\iint d\omega_\s d\omega_\i \,
    \phi(\omega_\s-\omega_\i)
	\adagger_\s(\omega_\s) \adagger_\i(\omega_\i)
	\vert \vac \rangle,   
    \label{eq:psi-twin}
\end{align}
where $\hat{a}_\s^\dagger(\omega)$ and $\hat{a}_\i^\dagger(\omega)$ are the creation operators for the signal and idler photons of frequency $\omega$, and the prefactor $\zeta$ represents the conversion efficiency of the SPDC process. The detailed derivation is presented in Appendix~A. In the phase-matching condition, $\phi(\omega)=\sinc(\omega T_\mathrm{e}/4)$, the parameter $T_\mathrm{e}$ denotes the so-called entanglement time \cite{Fei:1997es}. When the entanglement time is sufficiently long compared to timescales of the dynamics under investigation, i.e., $T_\mathrm{e} \to \infty$, the phase-matching function simplifies to $\phi(\omega_\mathrm{s}-\omega_\mathrm{i}) \simeq \delta(\omega_\mathrm{s}-\omega_\mathrm{i})$, leading to a simplified expression for the quantum state of the generated twin:
\begin{align}
	\lvert \psi_\twin \rangle
	=
    \zeta
	\int d\omega\,
	\adagger_\s(\omega) \adagger_\i(\omega)
	\vert \vac \rangle.
    \label{eq:psi-twin-simple}
\end{align}
This indicates that the frequency of the signal photon can be precisely reconstructed by measuring the frequency of the idler photon. Therefore, this study considered the limit $T_\mathrm{e} \to \infty$, as opposed to the CW pumping case wherein the limit $T_\mathrm{e}\to 0$ results in strong frequency correlations between the twin photons. The scenario wherein $T_\mathrm{e}$ takes a finite value is discussed in Appendix~C.

The free Hamiltonian of the radiation field is expressed as $\hat{H}_{\rm field}=\sum_{\sigma=\mathrm{s},\mathrm{F}}\int d\omega\,  \hbar\omega\,\hat{a}_{\sigma}^\dagger(\omega)\hat{a}_{\sigma}(\omega)$, where the operator $\hat{a}_\mathrm{FL}^\dagger(\omega)$ creates a spontaneously emitted photon of frequency $\omega$. The positive frequency component of the field operator is given by $\E_\sigma^+(t)=(2\pi)^{-1}\int d\omega\,\hat{a}_\sigma(\omega)e^{-i\omega t}$, whereas the negative frequency component is $\E_\sigma^-(t)= \E_{\sigma}^+(t)^\dagger$. We further assumed that the bandwidth of the fields is negligible compared to the central frequency \cite{loudon2000quantum}.

\section{Time- and frequency-resolved two-photon coincidence signal}
We investigated the time- and frequency-resolved two-photon coincidence signal, which facilitates the simultaneous acquisition of spectral and temporal information of photons through the combined use of the spectrometer and the DLD. To obtain this signal, the time-correlated single-photon counting (TCSPC) device measures the time correlation between the electric fields of the idler and fluorescence photons, both of which pass through the diffraction grating and subsequently the DLD before being output. 
However, the arrival time of photons measured by the DLD is affected by dead time and detector jitter, which introduces temporal blurring. The temporal distribution of the electric field of the detected photon is represented as the product of the temporal profile of the original electric field and the temporal distribution caused by the arrival time uncertainty of the DLD. Similarly, the measured photon frequency is influenced by the frequency resolution of the detector, leading to spectral blurring. The frequency distribution of the electric field of the detected photon can be described as the product of the frequency distribution of the original electric field and the frequency distribution associated with the uncertainty in the measured photon frequency.

To describe the information obtained at the DLD, we define the following two functions \cite{Dorfman:2016da,Gelin:2002ti,Yang:2023two}:
\begin{align}
	F_\mathrm{t}(t,\bar{t}_a)
	=
	\exp\left[-\frac{1}{2\sigma_\mathrm{t}^2} (t -\bar{t}_a)^2\right],
    \label{eq:time-gate function}	
\end{align}
\begin{align}
	F_\mathrm{f}(\omega,\bar\omega_a)
	=
    \frac{\sigma_\mathrm{f}}{i(\bar\omega_a - \omega) + \sigma_\mathrm{f}},
	\label{eq:frequency filter function 1}	
\end{align}
where $\bar\omega_a$ and $\bar{t}_a$ represent the center frequency and arrival time of photon~$a$ recorded at the DLD, respectively. Equation~\eqref{eq:time-gate function} describes the uncertainty of the arrival time of photon~$a$, $\bar{t}_a$, associated with the detection of photon~$a$ by the DLD.
Similarly, Eq.~\eqref{eq:frequency filter function 1} characterizes the uncertainty in the frequency of photon~$a$, $\bar\omega_a$, measured by the DLD.
The temporal resolution of the DLD is denoted by $\sigma_\mathrm{t}$, while $\sigma_\mathrm{f}$ represents the detector's frequency resolution, which is determined by both the number of grating grooves in the spectrometer and the position resolution of the DLD \footnote{Photons dispersed by the spectrometer arrive at different positions on the photocathode of the DLD based on their wavelengths. Enhancing the position resolution of the DLD improves the ability to distinguish closely spaced wavelengths in the photon spectrum, thereby increasing the frequency resolution. The position resolution of the DLD is primarily governed by the timing resolution of the electrical signals from the delay-line anode.}. Let $\hat{E}_{a}^{+}(t)$ denote the electric field before reaching the DLD. The information obtained at the DLDs can then be expressed as follows
\cite{Dorfman:2012no,Dorfman:2014bn,Dorfman:2016da}
\begin{align}
	\hat{\mathcal{E}}_a^+(\bar\omega_a,\bar{t}_a;t)
	=	
    \int_{-\infty}^\infty ds\,	
    F_{\rm f}(t-s,\bar\omega_a)
    F_\mathrm{t}(s,\bar{t}_a)
	\hat{E}_a^+(s),
	\label{eq:gated-electric-field}
\end{align}
where $F_\mathrm{f}(t,\bar\omega_a)$ is the Fourier transform of Eq.~\eqref{eq:frequency filter function 1}, i.e., $F_\mathrm{f}(t,\bar\omega_a) = \sigma_\mathrm{f} \theta(t) e^{-i \bar\omega_a t - \sigma_\mathrm{f}t}$ with $\theta(t)$ representing the Heaviside step function. The TCSPC device subsequently measures the time correlation between the electric fields detected at the two DLDs, $\hat{\mathcal{E}}_{\rm F}(\bar\omega_{\rm F},\bar{t}_{\rm F};t)$ and $\hat{\mathcal{E}}_{\rm i}(\bar\omega_{\rm i},\bar{t}_{\rm i};t)$, resulting in the time- and frequency-resolved two-photon coincidence signal \cite{Dorfman:2012no},
\begin{widetext}
\begin{align}
S(\bar\omega_{\rm F},\bar{t}_{\rm F};\bar\omega_{\i},\bar{t}_{\i})
	=
	\int_{-\infty}^\infty dt
	\int_{-\infty}^\infty  ds
	\,
	\tr\left[  
	   \hat{\mathcal{E}}_\mathrm{F}^{-}(\bar\omega_\mathrm{F},\bar{t}_\mathrm{F};t)
	   \hat{\mathcal{E}}_{\rm F}^{+}(\bar\omega_{\rm F},\bar{t}_{\rm F};t)
	   \hat{\mathcal{E}}_{\i}^{-}(\bar\omega_{\i},\bar{t}_{\i};s)
	   \hat{\mathcal{E}}_{\i}^{+}(\bar\omega_{\i},\bar{t}_{\i};s)
	   \hat{\rho}(t)     
    \right],      
	\label{eq:tpc-signal} 
\end{align}
\end{widetext}
where $\hat{\rho}(t)$ denotes the density operator for the total system, with $\hat{\rho}(-\infty) = \rho_\mathrm{mol}^\mathrm{eq} \otimes \lvert \psi_\twin \rangle\langle \psi_\twin \rvert$, and $\rho_\mathrm{mol}^\mathrm{eq}$ represents the thermal equilibrium state of the molecule's photoactive degrees of freedom. The term $\hat{\rho}(t)$ in Eq.~\eqref{eq:tpc-signal} can be perturbatively expanded with respect to the molecule-field interaction, $\hat{H}_{\rm mol-field}$, up to the fourth order. The signal consists of two contributions classified as rephasing and non-rephasing stimulated emission (SE) \cite{Mukamel:1995us}. Accordingly, Eq.~\eqref{eq:tpc-signal} is expressed as follows:
\begin{widetext}
\begin{multline}
	S(\bar\omega_{\rm F},\bar{t}_{\rm F};\bar\omega_{\i},\bar{t}_{\i})
	=
	\frac{\zeta^2}{2}
	\Re
	\int_{0}^\infty dt_3 \int_{0}^\infty dt_1 \,
	F_\mathrm{t}(t_3+\bar{t}_\mathrm{F},\bar{t}_\mathrm{F})
	F_\mathrm{t}(t_1+\bar{t}_{\i},\bar{t}_{\i}) \,
	e^{-(\sigma_{\rm f}-i\bar\omega_{\rm F})t_3}	
\\
	\times		
	\left[
	   e^{-(\sigma_\mathrm{f}+i\bar\omega_{\i})t_1}
	   \Phi_{\SE}^{\rephasing}(t_3,\bar{t}_{\rm F}+\bar{t}_{\i},t_1)
	   +
	   e^{-(\sigma_{\rm f}-i\bar\omega_{\i})t_1}
	   \Phi_{\SE}^{\nonrephasing}(t_3,\bar{t}_{\rm F}+\bar{t}_{\i},t_1)
	\right],
	\label{eq:tpc-signal-ultrafast-detection}	
\end{multline}
\end{widetext}
where $\Phi_{\SE}^{\rephasing}(t_3,t_2,t_1)$ and $\Phi_{\SE}^{\nonrephasing}(t_3,t_2,t_1)$ denote the rephasing and non-rephasing response functions, respectively. In deriving Eq.~\eqref{eq:tpc-signal-ultrafast-detection}, we assumed that the time resolution of the detector is sufficiently short compared to the system dynamics timescale, $\sigma_\mathrm{t} \to 0$. The derivation of this expression is provided in Appendix~C.

\section{Comparison of quantum and classical spectroscopy} 
The SE contribution to the absorptive 2D spectrum obtained using heterodyne-detected photon echo \cite{Brixner:2005wu} and phase-modulated fluorescence-detected 2D electronic spectroscopy \cite{Tekavec:2007ij,Lott:2011kd,Agathangelou:2021ph} in the impulsive limit is expressed as
\begin{align}
    \mathcal{S}_{\SE}(\omega_3,t_2,\omega_1)
    =
    \mathcal{S}_{\SE}^{\rephasing}(\omega_3,t_2,\omega_1)
    +
    \mathcal{S}_{\SE}^{\nonrephasing}(\omega_3,t_2,\omega_1),
    \label{eq:absorptive-2D-spectrum}	    
\end{align}
where $\mathcal{S}_{\SE}^{(x)}(\omega_3,t_2,\omega_1)$ represents the real part of the Fourier-Laplace transform of $\Phi_{\SE}^{(x)}(t_3,t_2,t_1)$. The signal in Eq.~\eqref{eq:tpc-signal-ultrafast-detection} is similar to the SE contribution of the absorptive 2D spectrum in Eq.~\eqref{eq:absorptive-2D-spectrum}, except that the signal in Eq.~\eqref{eq:tpc-signal-ultrafast-detection} depends on the shape of the function in Eq.~\eqref{eq:time-gate function}. When photon detectors with appropriate temporal and frequency resolution are used, the proposed quantum spectroscopy provides the following spectral information. The signal in Eq.~\eqref{eq:tpc-signal-ultrafast-detection} can temporally resolve the dynamics of the electronic excited state by recording $\bar{t}_\mathrm{F}+\bar{t}_\mathrm{i}$, that is, the sum of the detection times of fluorescence photons and heralded photons. Owing to the frequency correlations between the entangled photons, which satisfy the condition $\omega_\s=\omega_\i$, the proposed quantum spectroscopy also corresponds to the SE contribution of the absorptive 2D spectra. This is achieved through the measurement of the frequencies $\bar\omega_\mathrm{F}$ and $\bar\omega_{\i}$. By selectively detecting the SE contributions, the complexity of analyzing multidimensional spectra is expected to be reduced, thereby facilitating the extraction of information on the excited molecular state dynamics, as demonstrated in the numerical results.

Notably, a fundamental difference exists between the proposed quantum spectroscopy and conventional 2DES regarding the limits of time and frequency resolution. 
In 2DES, as described in Eq.~\eqref{eq:absorptive-2D-spectrum}, when all pulses are of very short duration, the resulting two-dimensional signal directly represents the molecule's third-order response function. In contrast, in the quantum spectroscopy, when the detector's time resolution is sufficiently short, the signal in Eq.~\eqref{eq:tpc-signal-ultrafast-detection} is given by the convolution of the third-order response function and the function in Eq.~\eqref{eq:time-gate function}. Under these conditions, the coherence between the electronic ground and excited states during $t_1$ and $t_3$ decays more rapidly than the original decay lifetime owing to the time profile of the function in Eq.~\eqref{eq:time-gate function}, thereby reducing the frequency resolution.
Indeed, in the case where the detector's time resolution is infinitely short, $F_\mathrm{t}(t,\bar{t}_a)=\delta(t-\bar{t}_a)$, Eq.~\eqref{eq:tpc-signal-ultrafast-detection} reduces to a time-resolved signal that has lost frequency resolution:
\begin{align}
	S(\bar\omega_\mathrm{F},\bar{t}_\mathrm{F};\bar\omega_{\i},\bar{t}_{\i})
	=
	\zeta^2	
	\Re	\,
    \Phi_{\SE}^{\rephasing}(0, \bar{t}_{\i} + \bar{t}_\mathrm{F} ,0).
	\label{eq:tpc-signal-no-frequency-resolution} 
\end{align}
Appendix~D details the numerical examination of the impact of the time resolution of the detector on the quantum spectroscopy. 

\begin{table}
    \begin{center}
    \caption{Single excitation Hamiltonian matrix elements in units of ${\rm cm}^{-1}$.}
    \begin{tabular*}{6cm}{@{\extracolsep{\fill}}cccc} \hline
    Pigment & 1     & 2     & 3     \\ \hline
    1       & 12500 & 30    & 30    \\ 
    2       & 30    & 12400 & $-40$ \\ 
    3       & 30    & $-40$ & 12200 \\ \hline
    \end{tabular*}
    \label{table1}
    \end{center}
\end{table}

\section{Numerical results} 
The advantages of the selective SE contributions are demonstrated through numerical calculations of a simple model system. We consider electronic excitations in a coupled trimer, described by the molecular Hamiltonian defined in Ref.~\citenum{Fujihashi:2023pr}. The third-order response functions in Eq.~\eqref{eq:tpc-signal-ultrafast-detection} are calculated using the second-order cumulant expansion with respect to the electronic energy fluctuations induced by the environment \cite{Zhang:1998eo}, and electronic energy transfer is modeled using the secular Redfield equation \cite{Redfield:1957tc}. The electronic excitation of each pigment is coupled to an environment characterized by the overdamped Brownian oscillator model \cite{Mukamel:1995us}. The timescale of environmental reorganization and the reorganization energy were set to $50\,{\rm fs}$ and $55\,{\rm cm}^{-1}$, respectively. The temperature was set to $77\,{\rm K}$. The excitation energy of the pigments and the interaction strength between the pigments are summarized in Table~\ref{table1}. For simplicity, the transition dipole moments of the pigments were assumed to be parallel and were set to the same value for all pigments.

\begin{figure}
    \centering
    \includegraphics{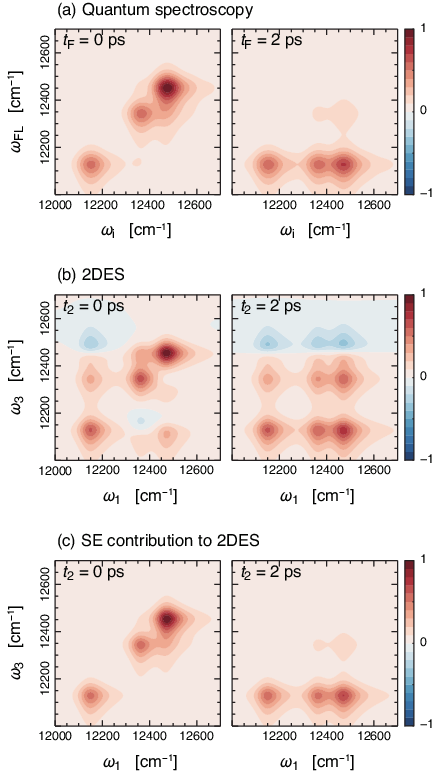}
    \caption{2D spectra obtained by the quantum spectroscopy described in Eq.~\eqref{eq:tpc-signal-ultrafast-detection}. (b) The absorptive 2D spectrum obtained using the classical pulses in the impulsive limit. (c) SE contribution to the absorptive 2D spectra. In panel (a), the detection time of the idler photon was set as $\bar{t}_\mathrm{i}=0$. The time resolution of the detector and the frequency resolution of the spectrometer were set to $\sigma_\mathrm{t}=400\,{\rm fs}$ and $\sigma_\mathrm{f}=0$, respectively. Normalization of contour plots~(a)--(c) is such that the maximum value of each spectrum at $\bar{t}_\mathrm{F}=0$ ($t_2=0$) is unity, and equally spaced contour levels ($0$, $\pm 0.1$, $\pm 0.2$, \dots) are shown.
    }
    \label{fig2}
\end{figure}

Figure~\ref{fig2}(a) shows the calculated 2D spectra obtained using the quantum spectroscopy described in Eq.~\eqref{eq:tpc-signal-ultrafast-detection}. The detection time of idler photons is fixed to $\bar{t}_{\i}=0$. The time resolution of the detector and the frequency resolution of the spectrometer are set to $\sigma_\mathrm{t}=400\,{\rm fs}$ and $\sigma_\mathrm{f}=0$, respectively. The detected times of the fluorescence photons are $\bar{t}_\mathrm{F}=0$ and $2\,{\rm ps}$. the absorptive 2D spectrum obtained using the photon echo technique in the impulsive limit is shown in Fig.~\ref{fig2}(b). Figure~\ref{fig2}(c) depicts the SE contribution to the absorptive 2D spectra presented in Fig.~\ref{fig2}(b). The spectrum at $t_2=0\,{\rm ps}$ in Fig.~\ref{fig2}(c) exhibits three diagonal peaks. Over time, energy transfer from higher-energy excitons to lower-energy excitons causes the disappearance of two diagonal peaks derived from high-energy excitons, along with the emergence of two cross-peaks. However, conventional 2D Fourier transform photon echo measurements observe only the sum of GSB, SE, and ESA contributions, making it impossible to isolate the SE contribution. Consequently, these measurements require careful interpretation of complex spectra. For example, the absorptive 2D spectrum in Fig.~\ref{fig2}(b) exhibits negative peaks resulting from the ESA, but the peak positions are shifted from their original ESA positions because of the overlap with nearby positive GSB peaks. This overlap hinders the extraction of accurate information regarding the excited state. In contrast, as shown in Fig.~\ref{fig2}(a), the signal obtained from the quantum spectroscopy contains spectral information equivalent to the SE contribution, except for the reduced frequency resolution caused by the detector's time resolution, $\sigma_\mathrm{t}$. The impact of the detector's time resolution is detailed in Appendix~D. Therefore, the proposed quantum spectroscopy presents a promising approach for accurately elucidating energy transfer processes in photosynthetic proteins with multiple pigments, which are often challenging to analyze using conventional 2D photon echo techniques. 

However, the time resolution of the DLD is on the order of several hundred picoseconds \cite{Iso:2024ca}, which is insufficient to observe excitation energy transfer in photosynthetic proteins \cite{Brixner:2005wu,Ishizaki:2012kf} and organic materials \cite{Kato:2018no} where subpicosecond time resolution is required. This limitation applies not only to the DLD but also to current single-photon imagers such as single-photon avalanche diode arrays \cite{Bruschini:2019si} and multi-anode single-photon imagers \cite{Oleksiievets:2022si}. One potential solution is to combine these detectors with streak tubes, which will enable time resolutions of approximately 200 fs.

\section{Conclusion} 
We demonstrated a quantum spectroscopy technique using a time-resolved fluorescence approach that is feasible with current photon detection technologies. The proposed quantum spectroscopy offers two significant advantages over conventional 2DES by leveraging the non-classical photon correlations of entangled photon pairs. First, the proposed quantum spectroscopy provides two-dimensional spectra without requiring the control of multiple pulsed lasers. Second, whereas conventional 2D spectra contain GSB, SE, and ESA contributions, the proposed spectroscopy selectively detects the SE contribution alone. This significantly reduces spectral complexity, facilitating straightforward extraction of information regarding the excited-state dynamics of molecular systems. These benefits are similar to those of previously proposed spectroscopic method \cite{Ishizaki:2020jl,Fujihashi:2024pa}. However, this approach achieves sufficient signal intensities that can be readily detected with current photon detection technologies such as the DLD, rendering it a practical method. Our findings are anticipated to pave way to the first experimental application of real-time observation of dynamic processes in molecular systems using quantum entangled photon pairs.

\begin{acknowledgments}
This study was supported by the MEXT Quantum Leap Flagship Program (Grant Number~JPMXS0118069242) and JSPS KAKENHI (Grant Number~JP21H01052). Y.F. acknowledges support from JSPS KAKENHI (Grant Number~JP23K03341).
\end{acknowledgments}

\appendix

\section{Entangled photon pairs}
\label{sec:appendix1}
\renewcommand{\theequation}{\ref{sec:appendix1}\arabic{equation}}
\setcounter{equation}{0}

This section details the derivation of Eq.~\eqref{eq:psi-twin}. For simplicity, we consider electric fields inside a one-dimensional nonlinear crystal of length $L$, subject to the degenerate type-II PDC process. In the weak down-conversion regime, the state of the generated twin photons is given by \cite{Grice:1997ht}
\begin{align}
	\lvert \psi_\twin \rangle
	=
	\int d\omega_1 \int d\omega_2
	\,
	f(\omega_1,\omega_2)
	\adagger_\s(\omega_1) \adagger_\i(\omega_2)
	\vert \vac \rangle,
	\label{eq:photon-state-SI}
\end{align}
where $\vert \vac \rangle$ denotes the photon vacuum state, and $\hat{a}_\s^\dagger(\omega)$ and $\hat{a}_\i^\dagger(\omega)$ are the creation operators of the signal and idler photons, respectively. The two-photon amplitude, $f(\omega_1,\omega_2)$, is expressed as 
\begin{align}
	f(\omega_1,\omega_2) 
	= 
	\zeta \alpha_\p(\omega_1+\omega_2) 
	\sinc\frac{\Delta k(\omega_1,\omega_2)L}{2},
	\label{eq:JSA-SI}    
\end{align}
where $\alpha_\p(\omega)$ is the normalized pump envelope, $\Delta k(\omega_1,\omega_2)$ represents the wave vector mismatch among the input and output photons, and $\zeta$ corresponds to the conversion efficiency of the PDC process. In the degenerate type-II PDC, $\Delta k(\omega_1,\omega_2)$ may be approximated linearly around the central frequencies of the two beams \cite{Keller:1997hj}, 
\begin{align}
	\Delta k(\omega_1,\omega_2) L = (\omega_1 - \omega_\p/2)T_\s + (\omega_2 - \omega_\p/2)T_\i
\end{align}
with $T_\lambda=(v_\p^{-1} - v_\lambda^{-1})L$. Here, $v_\p$, $v_\s$, and $v_\i$ are the group velocities of the pump laser with central frequency $\omega_\p$, signal beam, and idler beam, respectively. For simplicity, we assume the symmetric group-velocity matching, wherein the sum of the inverse group velocities of the signal and idler photons equals twice the pump inverse group velocity  ($T_\s=-T_\i$) \cite{Ansari:2018ta,Gatti:2018he}. This assumption leads to a two-photon amplitude characterized by the following form:
\begin{align}
	f(\omega_1,\omega_2)
	=
	\zeta \alpha_\p(\omega_1+\omega_2) 
	\phi(\omega_1 - \omega_2),
\end{align}
where $\phi(\omega)=\sinc[\omega T_\e/4]$. Here, the so-called entanglement time $T_\e= |T_\s - T_\i |$ is the maximum time difference between twin photons leaving the crystal \cite{Fei:1997es}. The analysis in this study is limited to the case of impulsive pumping in the PDC process, that is, $\alpha_\p(\omega)=1$. Consequently, we obtain the expression of the twin photon state as
\begin{align}
	\lvert \psi_\twin \rangle
	=
    \zeta
	\int d\omega_1 \int d\omega_2 \,
    \phi(\omega_1-\omega_2)
	\adagger_\s(\omega_1) \adagger_\i(\omega_2)
	\vert \vac \rangle.
\end{align}

\section{Hamiltonian}
\label{sec:appendix2}
\renewcommand{\theequation}{\ref{sec:appendix2}\arabic{equation}}
\setcounter{equation}{0}

We consider the Hamiltonian, 
\begin{align}
    \hat{H} = \hat{H}_{\rm mol} + \hat{H}_{\rm field} + \hat{H}_{\rm mol-field}.
\end{align}
The first term is the Hamiltonian to describe photoactive degrees of freedom in molecules. The second term, $\hat{H}_{\rm field}=\sum_{\sigma=\mathrm{s},\mathrm{F}}\int d\omega\,  \hbar\omega\,\hat{a}_{\sigma}^\dagger(\omega)\hat{a}_{\sigma}(\omega)$ describes the free radiation field,  where $\hat{a}_\mathrm{F}^\dagger(\omega)$ is the creation operator of a spontaneously emitted photon of frequency $\omega$. We employ the rotating-wave approximation for the molecule-field interaction as 
\begin{align}
    \hat{H}_{\rm mol-field}=-\hat{\mu}_+ \E_\s^+(t) -\hat{\mu}_+ \E_\mathrm{F}^+(t) + \mathrm{h.c.}, 
\end{align}
where $\hat{\mu}_+$ denotes the transition dipole operator to describe the optical transition from the ground state to the single excited state. The deexcitation from the single excited state to the ground state is described by $\hat{\mu}_{-}=\hat{\mu}_{+}^\dagger$. The positive frequency component of the field operator is given by $\E_\sigma^+(t)=(2\pi)^{-1}\int d\omega\,\hat{a}_\sigma(\omega)e^{-i\omega t}$, and the negative frequency component is $\E_\sigma^-(t)= \E_{\sigma}^+(t)^\dagger$. We further assume that the bandwidth of the fields is negligible compared to the central frequency \cite{loudon2000quantum}.

\section{Time- and frequency-resolved two-photon coincidence signal}
\label{sec:appendix3}
\renewcommand{\theequation}{\ref{sec:appendix3}\arabic{equation}}
\setcounter{equation}{0}

The time- and frequency-resolved two-photon coincidence signal in Fig.~1 is written as \cite{Dorfman:2012no},
\begin{widetext}
\begin{align}
S(\bar\omega_\mathrm{F},\bar{t}_\mathrm{F};\bar\omega_{\i},\bar{t}_{\i})
	=
	\int_{-\infty}^\infty dt
	\int_{-\infty}^\infty  ds
	\,
	\tr[
	\hat{\mathcal{E}}_\mathrm{F}^{-}(\bar\omega_\mathrm{F},\bar{t}_\mathrm{F};t)
	\hat{\mathcal{E}}_\mathrm{F}^{+}(\bar\omega_\mathrm{F},\bar{t}_\mathrm{F};t)
	\hat{\mathcal{E}}_{\i}^{-}(\bar\omega_{\i},\bar{t}_{\i};s)
	\hat{\mathcal{E}}_{\i}^{+}(\bar\omega_{\i},\bar{t}_{\i};s)
	\hat{\rho}(t)], 
	\label{eq:tpc-signal-SI}
\end{align}
where the density operator $\hat{\rho}(t)$ describes a state of the total system with the condition of $\hat{\rho}(-\infty)=\lvert 0 \rangle\langle 0 \rvert \otimes \lvert \psi_\twin \rangle\langle \psi_\twin \rvert$. By substituting Eq.~\eqref{eq:gated-electric-field}, we perform integration by parts on Eq.~\eqref{eq:tpc-signal-SI}. Consequently, we obtain
\begin{align}
	S(\bar\omega_\mathrm{F},\bar{t}_\mathrm{F};\bar\omega_{\i},\bar{t}_{\i})
	&=
	\frac{1}{2}
	\,
	\Re
	\int_{-\infty}^\infty dt
	\int_{-\infty}^{t} dt'
	\int_{-\infty}^\infty  ds
	\int_{-\infty}^{s}  ds'
	\,	
	F_\mathrm{t}(t,\bar{t}_\mathrm{F})
	F_\mathrm{t}(t',\bar{t}_\mathrm{F})	
	F_\mathrm{t}(s,\bar{t}_{\i})
	F_\mathrm{t}(s',\bar{t}_{\i})	
	e^{-(\sigma_\mathrm{f}-i\bar\omega_\mathrm{F})(t-t')}		
\notag \\
	&\quad\times
	\left(
	e^{-(\sigma_\mathrm{f}-i\bar\omega_\i)(s-s')}	
	\tr[
	\hat{E}_\mathrm{F}^{-}(t')
	\hat{E}_\mathrm{F}^{+}(t)
	\hat{E}_{\i}^{-}(s')
	\hat{E}_{\i}^{+}(s)
	\hat{\rho}(t)]
	\right.
\notag \\
	&
	\quad+
	\left.	
	e^{-(\sigma_\mathrm{f}+i\bar\omega_\i)(s-s')}		
	\tr[	
	\hat{E}_\mathrm{F}^{-}(t')
	\hat{E}_\mathrm{F}^{+}(t)
	\hat{E}_{\i}^{-}(s)
	\hat{E}_{\i}^{+}(s')
	\hat{\rho}(t)]
	\right).		
	\label{eq:tpc-signal2-SI}			
\end{align}	
The term $\hat{\rho}(t)$ in Eq.~\eqref{eq:tpc-signal2-SI} can be perturbatively expanded with respect to the molecule-field interaction, $\hat{H}_{\rm mol-field}$, up to fourth-order. The signal is expressed as the sum of two contributions, which are classified into rephasing and non-rephasing stimulated emission (SE). Consequently, Eq.~\eqref{eq:tpc-signal2-SI} can be expressed as
\begin{align}
	S(\bar\omega_\mathrm{F},\bar{t}_\mathrm{F};\bar\omega_{\i},\bar{t}_{\i})
	&=
	\frac{\zeta^2}{2}
	\,	
	\Re
	\int_{0}^\infty dt_3
	\int_{0}^\infty dt_2
	\int_{0}^\infty dt_1
	\int_{-\infty}^\infty  ds
	\,
	F_\mathrm{t}(t_3+t_2+t_1 -s,\bar{t}_\mathrm{F})
\notag \\
	&\quad\times
	F_\mathrm{t}(t_2 + t_1 -s,\bar{t}_\mathrm{F})	
	F_\mathrm{t}(s,\bar{t}_{\i})
	F_\mathrm{t}(s-t_1,\bar{t}_{\i})
	e^{-(\sigma_\mathrm{f}-i\bar\omega_\mathrm{F})t_3}	
	D(-s) D(t_1-s)
\notag \\
	&\quad\times		
	\left[
	e^{-(\sigma_\mathrm{f}+i\bar\omega_{\i})t_1}
	\Phi_{\SE}^{\rephasing}(t_3,t_2,t_1)
	+
	e^{-(\sigma_\mathrm{f}-i\bar\omega_{\i})t_1}
	\Phi_{\SE}^{\nonrephasing}(t_3,t_2,t_1)
	\right],
	\label{eq:tpc-signal3}	
\end{align}
where $D(t)$ is the Fourier transform of the phase matching function defined by
\begin{align}
    D(t)
    =
    (2\pi)^{-1}\int_{-\infty}^\infty d\omega\, \phi(\omega) e^{-i\omega t}.
    \label{eq:PMF-Fourier}    
\end{align}
We assume that the time resolution of the detector is sufficiently short compared to the timescale of the system dynamics.
By letting $F_\mathrm{t}(t) \to \delta(t)$ for two of the four functions $F_\mathrm{t}(t)$, Eq.~\eqref{eq:tpc-signal3} simplifies to
\begin{align}
	S(\bar\omega_\mathrm{F},\bar{t}_\mathrm{F};\bar\omega_{\i},\bar{t}_{\i})
	&=
	\frac{\zeta^2}{2}
	\,	
	\Re
	\int_{0}^\infty dt_3
	\int_{0}^\infty dt_1
	\,
	F_\mathrm{t}(t_3+\bar{t}_\mathrm{F},\bar{t}_\mathrm{F})
	F_\mathrm{t}(t_1+\bar{t}_{\i},\bar{t}_{\i})
	e^{-(\sigma_\mathrm{f}-i\bar\omega_\mathrm{FL})t_3}	
	D(-t_1-\bar{t}_{\i}) D(-\bar{t}_{\i})
\notag \\
	&\quad\times		
	\left[
	e^{-(\sigma_\mathrm{f}+i\bar\omega_{\i})t_1}
	\Phi_{\SE}^{\rephasing}(t_3,\bar{t}_{\i}+\bar{t}_\mathrm{F},t_1)
	+
	e^{-(\sigma_\mathrm{f}-i\bar\omega_{\i})t_1}
	\Phi_{\SE}^{\nonrephasing}(t_3,\bar{t}_{\i}+\bar{t}_\mathrm{F},t_1)
	\right].
	\label{eq:tpc-signal4}	
\end{align}
\end{widetext}
When the entanglement time is sufficiently long compared to timescales of dynamics under investigation, namely $T_\mathrm{e} \to \infty$, Eq.~\eqref{eq:PMF-Fourier} simplifies to $D(t)=\delta(t)$, resulting in Eq.~\eqref{eq:tpc-signal-ultrafast-detection}.

\begin{figure}
    \centering
    \includegraphics{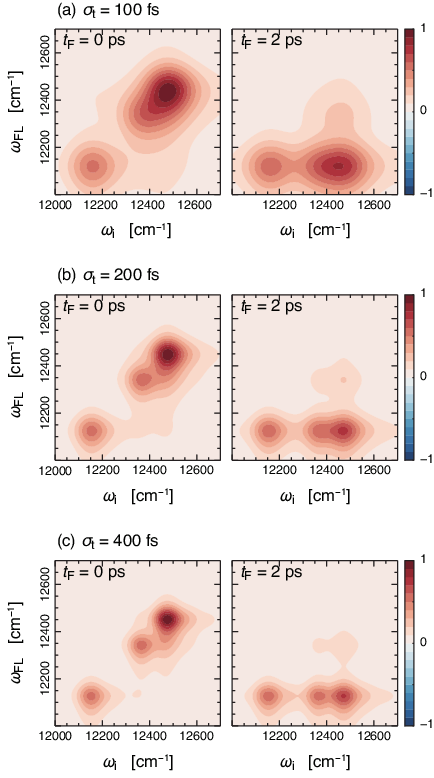}
    \caption{(a) Impact of the time resolution of the detector on the quantum spectroscopy described in Eq.~\eqref{eq:tpc-signal4}. The detection time of the idler photon is fixed to $\bar{t}_\mathrm{i}=0$. The entanglement time is set to $T_\mathrm{e}=1600\,{\rm fs}$. Other parameters are the same as those in Fig.~\ref{fig2}. Normalization of contour plots~(a)--(c) ensures that the maximum value of each spectrum at $\bar{t}_\mathrm{F}=0$ is unity, and equally spaced contour levels ($0$, $\pm 0.1$, $\pm 0.2$, \dots) are shown.
    }
    \label{figS1}
\end{figure}

\begin{figure}
    \centering
    \includegraphics{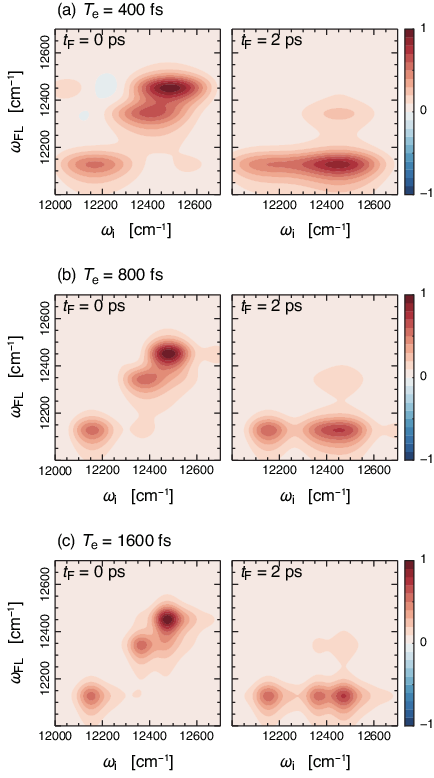}
    \caption{Impact of the entanglement time on the quantum spectroscopy described in Eq.~\eqref{eq:tpc-signal4}. The detection time of the idler photon is fixed to $\bar{t}_\mathrm{i}=0$. Other parameters are the same as those in Fig.~\ref{fig2}. Normalization of contour plots~(a)--(c) ensures that the maximum value of each spectrum at $\bar{t}_\mathrm{F}=0$ is unity, and equally spaced contour levels ($0$, $\pm 0.1$, $\pm 0.2$, \dots) are shown.
    }
    \label{figS2}
\end{figure}

\begin{figure}
    \centering
    \includegraphics{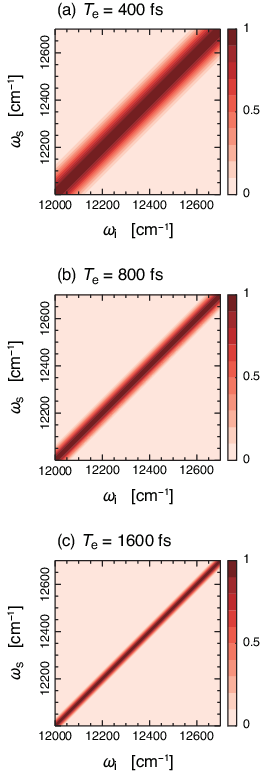}
    \caption{Two-photon spectral intensity, $|f(\omega_\s,\omega_\i)|^2$, in Eq.~\eqref{eq:JSA-SI} for (a) $T_\mathrm{e}=400\,{\rm fs}$, (b) $T_\mathrm{e}=800\,{\rm fs}$, and (c) $T_\mathrm{e}=1600\,{\rm fs}$. The pump laser is assumed to be impulsive, i.e., $\alpha_\p(\omega)=1$. Normalization of contour plots~(a)--(c) ensures that the maximum value of each spectrum is unity, and equally spaced contour levels ($0$, $0.1$, $0.2$, \dots) are shown.
    }
    \label{figS3}
\end{figure}

\section{Additional numerical results}
\label{sec:appendix4}
\renewcommand{\theequation}{\ref{sec:appendix4}\arabic{equation}}
\setcounter{equation}{0}

To supplement the discussion in in the main text,
we present additional numerical results in this section. The quantum spectroscopy calculations were performed using Eq.~\eqref{eq:tpc-signal4} instead of Eq.~\eqref{eq:tpc-signal-ultrafast-detection} to account for the effects of finite entanglement time (i.e., the phase-matching function).

\subsection{Impact of the time resolution of the detector on the 2D spectra}
We explore the impact of the time resolution of the detector on the 2D spectra. Figure~\ref{figS1} presents 2D spectra obtained with quantum spectroscopy for various values of $\sigma_\mathrm{t}$. The entanglement time is set to $T_\mathrm{e}=1600\,{\rm fs}$. All other parameters in Fig.~\ref{figS1} are consistent with those in Fig.~\ref{fig2}. As the time resolution of the DLD increases, the frequency resolution of the 2D spectra in Fig.~\ref{figS1} diminishes, as indicated by Eq.~\eqref{eq:tpc-signal-no-frequency-resolution} . For instance, at $\sigma_\mathrm{t}=100\,{\rm fs}$, the peaks corresponding to the two exciton states overlap because of the reduced frequency resolution. This overlap makes it challenging to extract detailed spectral information regarding these exciton states.

\subsection{Impact of the entanglement time on the 2D spectra}

Figure~\ref{figS2} illustrates 2D spectra obtained with quantum spectroscopy for varying values of the entanglement time. The detector's time resolution is set to $\sigma_\mathrm{t}=400\,{\rm fs}$. Other parameters in Fig.~\ref{figS2} are consistent with those in Fig.~\ref{fig2}. While the frequency resolution along the fluorescence photon's frequency axis remains unaffected by changes in entanglement time, shorter entanglement times degrade the frequency resolution along the idler photon's frequency axis. This degradation occurs because shorter entanglement times lead to a broader frequency distribution for the entangled photon pair. 

To clarify this effect, we analyze the two-photon frequency distribution of the entangled photon pair. Figure~\ref{figS3} depicts the two-photon spectral intensity, $|f(\omega_\s,\omega_\i)|^2$, as defined in Eq.~\eqref{eq:JSA-SI} for three cases: (a) $T_\mathrm{e}=400\,{\rm fs}$, (b) $T_\mathrm{e}=800\,{\rm fs}$, and (c) $T_\mathrm{e}=1600\,{\rm fs}$.
As shown in Fig.~\ref{figS3}(c), for $T_\mathrm{e}=1600\,{\rm fs}$, the frequencies $\omega_\s$ and $\omega_\i$ exhibit strong correlations, and the two-photon spectral intensity is concentrated near $\omega_\s = \omega_\i$. Consequently, the idler photon's frequency directly corresponds to the signal photon's frequency absorbed by the pigments. However, as the entanglement time decreases, the two-photon spectral intensity broadens along the anti-diagonal direction. For $T_\mathrm{e}=400\,{\rm fs}$, the signal photon's frequency paired with the detected idler photon is blurred by approximately $142\,{\rm cm}^{-1}$, as defined by the full width at half maximum of the two-photon spectral intensity. As a result, as shown in Fig.~\ref{figS2}(a), the frequency resolution along the idler photon's frequency axis decreases.
\bigskip

%

\end{document}